\documentclass[sigconf]{acmart}

\usepackage{array}
\usepackage{tabularx}
\usepackage{xcolor}

\AtBeginDocument{%
  }

\copyrightyear{2026}
\acmYear{2026}
\setcopyright{cc}
\setcctype{by}
\acmConference[ICAIF '26]{7th ACM International Conference on AI in Finance}{November 14--17, 2026}{Milan, Italy}
\acmBooktitle{7th ACM International Conference on AI in Finance (ICAIF '26), November 14--17, 2026, Milan, Italy}
\acmDOI{10.1145/3858063.3858739}
\acmISBN{979-8-4007-3087-0/2026/11}

\begin{document}

\title[More Mentions, Less Information?]{More Mentions, Less Information? Diagnosing Identity and Alignment in Reddit Embeddings for Volatility Forecasting}

\author{Zhiyu Zheng}
\affiliation{%
  \institution{University of Chicago} \city{Chicago} \state{Illinois} \country{USA} }
\email{zhengzhiyu6689@uchicago.edu}
\renewcommand{\shortauthors}{Zheng}

\begin{abstract}
Since the 2021 GameStop short squeeze, r/WallStreetBets (WSB) has been associated with episodes of extreme volatility, and its text is increasingly used to forecast volatility. A model using only large language model (LLM) embeddings of WSB posts attains an out-of-sample $R^2$ of $0.13$ for log realized variance. That number shows predictive power, but it does not reveal where the information comes from or how much the text adds beyond a firm's own volatility history. We answer these questions with a diagnostic protocol that nests embeddings on a heterogeneous autoregressive (HAR) benchmark, compares them with text-free firm-identity controls, shuffles the day-to-text pairing within firm-years to test day-specific alignment, and checks whether gains persist across periods and coverage groups. We apply the protocol to frozen LLM embeddings in a panel of $202{,}145$ mentioned stock-days, using retrospective rolling out-of-sample tests over 2015--2022. The results indicate that most of the standalone fit reflects firm or firm-year information: on firms seen in training, a text-free firm baseline reaches $R^2=0.32$, shuffling preserves $89$--$92\%$ of the standalone fit, and masking ticker strings reduces standalone $R^2$ to $-0.005$. Adding embeddings to HAR reduces equal-date MSE by $1.22\%$ ($95\%$ CI $[0.94,1.49]$). This small but statistically significant gain survives firm controls and ticker masking, but is largely eliminated by alignment shuffling. The gain is concentrated in 2015--2019 and is not distinguishable from zero in 2020--2022. Expanding coverage by $48\%$ through off-hours and weekend posts adds stock-days with no reliable positive increment. More mentions therefore do not imply more incremental forecasting information.
\end{abstract}



\maketitle

\section{Introduction}
\label{sec:intro}

A growing literature uses text representations, increasingly LLM embeddings, to predict financial outcomes such as returns~\citep{chen2024expected} and volatility~\citep{kogan2009predicting,atkins2018financial,rahimikia2021realised}, and these models often report high standalone accuracy. Yet an embedding of a security's social-media discussion encodes not only content specific to the day but also which firm is being discussed, its sector and discussion style, and broader firm-year conditions. A high out-of-sample $R^2$ for volatility may therefore reflect firm identity that a far simpler control would capture, rather than genuine information from the day's text. Diagnosing what drives this predictive power is therefore crucial for forecasting financial markets with embeddings.

Rather than asking whether an embedding model beats a benchmark, we introduce a protocol that asks where its predictive power comes from and separates three sources: volatility history, firm identity, and day-specific information. Volatility history is controlled by adding the embedding to the HAR model of~\citet{corsi2009simple}, so that we measure only what the embedding adds beyond past volatility. For firm identity, we compare the embeddings with baselines that know only which firm is being discussed, and we check whether the gain over HAR remains after adding a firm effect. For day-specific information, we shuffle which day's posts are matched to which day's volatility within each firm-year. The shuffle preserves firm-year grouping while disrupting the original day-to-text pairing. Incremental comparisons use nested models and equal-date loss aggregation, with confidence intervals obtained from a moving-block bootstrap.

We apply the protocol to Llama-3.1-8B embeddings of Reddit r/WallStreetBets (WSB) posts from 2012 to 2022, aggregated to $202{,}145$ mentioned stock-days, and evaluate the forecasts out of sample in rolling windows with test years 2015--2022. An embedding-only model attains an out-of-sample $R^2$ of $0.13$, but a baseline that knows only which firm is being forecast reaches $0.32$ on previously seen firms, and the shuffle leaves the standalone fit largely intact: most standalone predictability reflects which firm is being discussed. Added to HAR, embeddings reduce equal-date MSE by 1.22\%, a small but statistically significant improvement that survives firm controls. Yet this increment disappears under the shuffle: it requires the correct pairing of each day's posts with that day's outcome. This dependence shows that the gain comes from day-specific content, but it does not by itself establish that the posts anticipate the outcome: a post that merely describes an event of the day, such as an earnings release or a large price move, is already informative about the next session's volatility, and our design cannot tell such contemporaneous descriptions apart from forward-looking information. The increment is also fragile: statistically significant in 2015--2019, it is indistinguishable from zero in 2020--2022. Expanding the panel by $48\%$ with off-hours and weekend content brings no additional forecasting value: the newly covered stock-days contribute zero or negative increments.

Our contributions are threefold:
\begin{enumerate}
\item \textbf{A diagnostic protocol} that asks where a text embedding's predictive power comes from. It separates three sources, volatility history, firm identity, and day-specific information, by nesting the embedding on HAR, comparing it with text-free firm baselines, and shuffling the day-to-text pairing within each firm-year. The protocol is model-agnostic.

\item \textbf{Evidence that standalone accuracy overstates the text's forecasting ability.} On $202{,}145$ WSB stock-days, an embedding only model attains $R^2 = 0.13$, but most of this fit reflects firm identity, not day-specific content. The remaining increment over HAR, a 1.22\% reduction in MSE, vanishes once the day-to-text pairing is shuffled and is statistically significant in 2015--2019 but indistinguishable from zero in 2020--2022.

\item \textbf{More mentions, less information.} Recovering off-hours and weekend content enlarges the panel by $48\%$, yet the added stock-days carry no positive incremental value. The added stock-days change the composition of the panel without improving the forecasts.
\end{enumerate}

\section{Related Work}
\label{sec:related}
We build on research on investor attention and social media, text representations in financial prediction, and realized volatility forecasting. Our diagnostic approach also draws on work on probing neural representations.

\paragraph{Investor attention and social media.}
Retail attention predicts short-horizon price pressure~\citep{da2011search}, and social media provide a setting in which to study how investors share information and respond to it. Studies link investor disagreement to trading volume~\citep{cookson2020w}, Twitter mood to index movements~\citep{bollen2011twitter}, and sentiment around attention peaks to abnormal returns~\citep{ranco2015effects}. On WallStreetBets, due diligence recommendations predicted returns before the GameStop squeeze, but this predictability disappeared afterwards~\citep{bradley2024place}. The meme stock episode also highlighted the market impact of retail trading coordinated through social media~\citep{aggarwal2024meme}. Methods for measuring sentiment include finance dictionaries~\citep{loughran2011liability}, social media lexicons~\citep{HuttoGilbert2014}, and financial transformers~\citep{huang2023finbert}. We retain contextual LLM embeddings rather than reducing text to a sentiment score, and ask what information in these representations contributes to volatility forecasts.

\paragraph{Text representations in financial prediction.}
Text provides a high-dimensional source of economic information~\citep{GentzkowKellyTaddy2019TextAsData}. News text helps track and forecast macroeconomic activity~\citep{bybee2024business}, and LLM embeddings of news improve return prediction relative to simpler text representations~\citep{chen2024expected}. Combining sentiment with deep learning can also improve index volatility forecasts, although the gains vary across markets~\citep{ncume2022volatility}. These findings establish the predictive relevance of financial text, but predictive fit alone does not support a causal or structural interpretation~\citep{mullainathan2017machine,clayton2025faustian}. We examine which information drives the fit: how much reflects persistent firm information, how much depends on the correct day-to-text pairing, and what text adds beyond volatility history. We also test whether the measured increment persists across periods and changes in sample coverage.

\paragraph{Volatility forecasting.}
Realized measures aggregate intraday returns to provide less noisy proxies for daily volatility than squared daily returns~\citep{andersen2003modelling}, with foundations in the theory of quadratic variation~\citep{barndorffnielsen2002}. The HAR model of~\citet{corsi2009simple} is a widely used benchmark for realized volatility forecasting. Related approaches include MIDAS and Spline-GARCH~\citep{ghysels2006predicting,engle2008spline}, while comparisons of GARCH models illustrate the importance of evaluating forecasts against simple benchmarks~\citep{hansen2005forecast}. Text-based measures also capture policy uncertainty~\citep{baker2016measuring} and disaster concerns~\citep{manela2017news}. Closely related work by~\citet{rahimikia2021realised} uses financial news word embeddings to forecast stock-level realized volatility and evaluates their combination with HAR-family forecasts. We study retail forum text using contextual LLM embeddings, with a focus on firm identity and day-specific alignment as sources of predictive performance.

\paragraph{Probing representations.}
Linear probes test which information can be recovered from a frozen neural representation~\citep{alain2017understanding}. Their accuracy must be interpreted with care: a successful probe need not show that the underlying model uses the information, and controls are needed to distinguish information in the representation from patterns learned by the probe~\citep{belinkov2022probing}. Our linear forecasting head shares this logic. We adapt it to financial panels, using text-free firm baselines and within-firm-year shuffles to assess how much predictive performance depends on firm identity and day-specific alignment (Section~\ref{sec:method}).

\section{Data and Measurement}
\label{sec:data}

\subsection{Reddit Text and Embeddings}
\label{sec:data-text}

We study posts submitted to the Reddit WallStreetBets (WSB) subreddit between 2012 and 2022, obtained from the Pushshift Reddit archive~\citep{baumgartner2020pushshift}. Posts are matched to ticker symbols appearing in the title or body, a post mentioning several securities contributes one row per mentioned ticker, and each ticker string is mapped to the CRSP permanent identifier (\texttt{permno}) valid on the forecast date to which the post is assigned, using the CRSP name history. Matching capitalized strings also picks up ordinary words that happen to be tickers, such as DD (due diligence), YOLO or ON. We use how often a string is written as a cashtag, with a leading dollar sign, as a heuristic to flag such strings: \$GME appears as a cashtag in $16\%$ of its post rows and \$TSLA in $27\%$, whereas DD is written as a cashtag in only $0.1\%$ of its post rows. A string that appears in at least $50$ post rows and is written as a cashtag in fewer than $5\%$ of them is treated as a common word and counts as a mention only when written as a cashtag. Cashtag shares are computed on the session-window post rows before the rule is applied, so the classification uses text only, never an outcome. Of the $6{,}957$ strings that link session-window posts to labeled stock-days, $271$ are classified as common words in this way. Table~\ref{tab:link} compares the date-valid link with a static ticker table that keeps the earliest \texttt{permno} of each string, and reports sensitivity to the threshold. Each unique title-body text is embedded once with the frozen Llama-3.1-8B base model~\cite{llama31}, loaded in 8-bit precision. The title and body are concatenated, separated by a blank line, and empty or placeholder title or body fields are omitted. To keep embedding nearly one million texts tractable, inputs are truncated to at most $192$ tokens, and approximately $2{,}000$ of the longest posts were capped at $160$ tokens to fit GPU memory. In the main panel, $4.8\%$ of unique texts exceed the $192$-token cap, and the median text length is $17$ tokens. Section~\ref{sec:res-robust} shows that raising the cap to $4{,}096$ tokens does not raise the increment. We apply attention-masked mean pooling to the final-layer (layer-32) token representations and $L_2$-normalize each vector, yielding one $4{,}096$-dimensional embedding per unique text.

Reposting is common on the platform, and deduplication operates on post content. Embeddings are keyed to unique title-body texts, so exact reposts share a single text key. An exact-record filter removes rows that are identical on (text key, timestamp, \texttt{permno}), and within each stock-day we retain one embedding per unique text. Identical reposts are therefore counted once per stock-day, whereas near-duplicates that differ in wording remain. A stock-day embedding is the mean of these per-text embeddings for that (\texttt{permno}, forecast date) pair.

Our unit of analysis is a \emph{mentioned stock-day}: a (\texttt{permno}, forecast date) pair to which at least one WSB post is assigned under the applicable information-window rule (Section~\ref{sec:data-window}). The sample is conditional on both WSB mentions and label availability. Because forum attention is not random, the volatility characteristics of this sample may differ from those of the broader market. We estimate HAR and evaluate all textual increments within this sample. Our findings therefore concern mentioned stock-days, not the full equity cross-section, and do not imply trading profitability.

\subsection{Realized Volatility and the HAR Target}
\label{sec:data-rv}
We use realized-variance labels based on intraday TAQ prices, following established procedures in the realized-volatility literature~\cite{andersen2003modelling,dai2024}: prices are aggregated into 15-minute log returns ($26$ intraday intervals per day plus an overnight bin), and realized variance is the sum of squared returns, which we annualize and log-transform. The overnight return is defined as $r^{\text{on}}_t = \log(1 + R^{\text{CRSP}}_t) - \log(P^{\text{close}}_t / P^{\text{open}}_t)$, where $R^{\text{CRSP}}_t$ is the CRSP close-to-close return and the prices are the TAQ open and close. It therefore covers the previous close to the current open and inherits CRSP's adjustment for dividends and splits. Our forecasting benchmark is the Heterogeneous Autoregressive (HAR) model of Corsi~\cite{corsi2009simple}: the features \textsf{RV}, \textsf{RVw}, \textsf{RVm}, and \textsf{RVq} are trailing averages of log realized variance over $1$, $5$, $21$, and $63$ trading days, and the forecast target $Y_1$ is the log realized variance of the next eligible observation.

A stock-day enters the panel only if it passes standard cleaning filters: at least two nonzero intraday returns and positive RV, at least twelve trades, not a half-day trading session, and not an isolated log-RV spike or drop relative to neighboring days. The source RV panel further requires each security to have at least 150 available observations over the panel's full history.

We estimate all models on log realized variance and use mean squared error on this scale as the primary evaluation loss. As a secondary sensitivity check, we report QLIKE on variance levels~\citep{patton2011volatility}, using exponentiated log forecasts without bias correction.

\subsection{Information Window}
\label{sec:data-window}
We compare two panels constructed using different information windows. The \emph{same-day panel} uses only posts submitted on the forecast date before 16:00~ET. It excludes after-hours, weekend, and holiday posts. The \emph{session-window panel}, our main panel, assigns each post to the first official market-session close strictly after its timestamp ($13$:$00$~ET on early-close days and $16$:$00$~ET otherwise). Its information window runs from the previous session close up to, but not including, the forecast-date close. Under either rule, a stock-day is retained only if it has at least one eligible post and complete labels in the source RV panel.

The session-window rule recovers the discarded off-hours content and enlarges the panel by $48\%$ (Table~\ref{tab:panel}), and no stock-day of the same-day panel is lost. For $99.1\%$ of session-window stock-days the target is the next trading session. The exceptions arise when the observation for that session is dropped by the filters in Section~\ref{sec:data-rv}. We define a consecutive-session sample, the stock-days for which the target is the next session, as a sensitivity check. The target is the next eligible RV observation, which need not correspond to the immediately following trading session. The newly covered stock-days are spread across all sample years roughly in proportion to posting activity. The forum itself, however, grows sharply over time: the number of distinct securities mentioned per trading day rises roughly $3.1$-fold, from about $56$ per day in 2015--2019 to about $173$ in 2020--2022, peaking at about $262$ in 2021, a fact we return to in Section~\ref{sec:results}.

\begin{table}[t]
\caption{Panel sizes and test-sample composition. A stock-day is a security--forecast-date pair. OOS test counts cover 2015--2022. Within the session-window test sample, previously covered stock-days also appear in the same-day panel, whereas newly covered stock-days appear only in the session-window panel. The lower panel reports annual test counts for the session-window panel.}
  \label{tab:panel}
  \small
  \begin{tabular}{lrr}
    \toprule
    Sample & Stock-days & OOS test \\
    \midrule
    Same-day panel & 136{,}333 & 134{,}237 \\
    Session-window panel (main) & 202{,}145 & 198{,}564 \\
    \quad Consecutive-session subsample & 200{,}381 & 196{,}866 \\
    \midrule
    \multicolumn{3}{l}{Session-window test-sample partition} \\
    \quad Previously covered & --- & 134{,}237 \\
    \quad Newly covered & --- & \phantom{0}64{,}327 \\
    \bottomrule
  \end{tabular}

  \vspace{2pt}
  {\footnotesize\setlength{\tabcolsep}{3pt}
  \begin{tabular}{lrrrrrrrr}
    \toprule
    Year & 2015 & 2016 & 2017 & 2018 & 2019 & 2020 & 2021 & 2022 \\
    \midrule
Test stock-days & 5,549 & 12,346 & 15,571 & 16,892 & 18,540 & 39,042 & 65,747 & 24,877 \\
    \bottomrule
  \end{tabular}}
\end{table}

\section{Evaluation Design}
\label{sec:method}

This section describes the rolling design, the estimand and inference, and the comparisons that separate volatility history, firm identity and day-specific alignment.

\subsection{Rolling Out-of-Sample Forecasts}
\label{sec:method-rolling}

We use a 2--1--1 rolling scheme over 2012--2022: each rolling window trains on two years, selects hyperparameters on the next year, and tests on the following one, giving eight non-overlapping test years (2015--2022). We use a fixed two-year training window to accommodate potential changes in forum composition while maintaining a consistent estimation scheme across test years. Within each rolling window, we estimate common HAR coefficients by pooled OLS over all training stock-days, which provides a common benchmark across firms with uneven training coverage.

We evaluate embeddings through a nested extension of HAR. Embeddings are row-normalized to unit length and then standardized using training-window column means and standard deviations. The target is centered using its training-window mean. Let $Z$ denote the resulting $4{,}096$-dimensional stock-day embedding. Within each rolling window, we fit a ridge regression of the HAR training residuals on $Z$ and add its predictions to the fixed HAR forecasts. We select the ridge penalty by validation MSE from $24$ values log-spaced from $10^{-3}$ to $10^{10}$, each multiplied by the number of training stock-days. The candidate set also includes the unaugmented HAR forecast, so validation can reject the embedding component, although out-of-sample underperformance remains possible. The embeddings enter the regression directly, without dimension reduction. HAR coefficients remain fixed throughout this second stage, so the comparison measures the predictive value of adding embeddings without refitting the benchmark. Post counts enter in the same way through ordinary least squares, and firm effects through the shrinkage estimator of Section~\ref{sec:method-firm}, each fitted on the residual of the model it extends.

Our evaluation is retrospective. Llama 3.1 was pretrained using data extending beyond the end of our sample, so the rolling splits enforce out-of-sample evaluation of the forecasting head, not temporal separation of the encoder's pretraining data. The alignment shuffle tests dependence on day-specific text pairing but cannot rule out pretraining contamination. The Llama-2 comparison in Section~\ref{sec:res-robust} tests sensitivity to the encoder rather than resolving this limitation.

\subsection{Estimand and Inference}
\label{sec:method-estimand}
We summarize accuracy by the relative loss improvement of an augmented model over HAR,
\begin{equation}
  \Delta \;=\; \frac{L_{\mathrm{HAR}} - L_{\mathrm{AUG}}}{L_{\mathrm{HAR}}},
  \label{eq:delta}
\end{equation}
where $L$ denotes loss aggregated over the out-of-sample observations under the stated estimand. The primary loss is MSE on log-RV and the secondary loss is QLIKE. We form the ratio in Eq.~\eqref{eq:delta} only \emph{after} aggregating losses.

Because trading days differ greatly in the number of securities mentioned, we use an \emph{equal-date} estimand: We first average forecast losses across securities on each test date, then average these daily losses across test dates. Each date therefore receives equal weight, regardless of the number of securities mentioned. We report the pooled stock-day estimand as a robustness check.

For comparisons, we also report pooled out-of-sample $R^2$, defined as $1-\sum_i (y_i-\hat y_i)^2/\sum_i (y_i-\bar y)^2$. Here, $i$ indexes the test stock-days in the reported sample and $\bar y$ is their mean outcome. Within each reported sample, all models use the same denominator.

We construct confidence intervals using a circular moving-block bootstrap, resampling blocks of consecutive test dates separately within each test year. Each sampled date includes all securities observed on that date, preserving their cross-sectional dependence. We use five-trading-day blocks for the main analysis and block lengths of $1$, $10$, and $22$ trading days for sensitivity checks. We resample the out-of-sample observations without re-estimating the models or recomputing their predictions, so the confidence intervals reflect uncertainty in the evaluation losses rather than uncertainty from re-estimating the models. We report pointwise confidence intervals and permutation $p$-values without adjustment for multiple comparisons.

\subsection{Standalone Versus Incremental Predictability}
\label{sec:method-standalone}
A recurring pitfall in text-based asset pricing is to rely solely on a text signal's standalone predictive accuracy as evidence that it provides information beyond existing forecasting benchmarks. Standalone out-of-sample predictive accuracy measures how well an embedding-only model predicts the outcome, but it does not establish whether text adds predictive value beyond an existing benchmark. Volatility differs substantially across firms, and embeddings can encode persistent firm characteristics such as sector, products, and discussion style. An embedding-only model may therefore achieve a high $R^2$ without relying strongly on day-specific text. We distinguish this standalone fit from the reduction in forecast loss obtained by adding embeddings to HAR. The latter measures incremental predictive value relative to the specified benchmark.

\subsection{Firm-Identity Controls}
\label{sec:method-firm}
We use firm effects to control for persistent firm differences and a within-firm-year permutation test to assess dependence on day-specific alignment. The firm effect is a regularized one-hot encoding of \texttt{permno}: Within each rolling window, we compute the HAR residual for each training observation as the observed log realized variance minus the fitted HAR value. We then average these residuals within each firm and shrink the resulting firm-specific mean toward zero to obtain the firm effect. We select the shrinkage parameter by validation MSE from $20$ values log-spaced from $10^{-1}$ to $10^{4}$. We also allow validation to select the HAR forecast without any firm adjustment. Firms absent from the training sample receive a zero firm effect, so their HAR+firm forecasts equal the HAR forecasts. We call a test stock-day \emph{seen} when its firm appears in the training sample of the same rolling window and \emph{unseen} otherwise, and report results for all test stock-days and for the two scopes separately. Seen stock-days are 74.3\% of the test sample. We also report a standalone firm-mean baseline without HAR predictors or shrinkage. In each rolling window, it predicts log realized variance using the firm's training-sample mean. For firms absent from training, it uses the mean across all training observations. The one-hot baseline in Table~\ref{tab:firm} applies the same shrinkage estimator to deviations of log realized variance from its training mean, without HAR predictors. We then compare HAR${+}$firm with HAR${+}$firm${+}$embedding: if the embedding still improves the forecast once the firm effect is present, the increment is not explained by firm identity.

The permutation test uses a \emph{within-firm-year shuffle}. For each firm and calendar year, we randomly reassign its daily embeddings across its observed stock-days, leaving the outcomes and HAR predictors unchanged. We perform this reassignment separately within the training, validation, and test sets, so no embedding moves between these sets. The shuffle preserves the firm-year grouping but disrupts the original pairing between each day's text and its forecast target. We repeat the shuffle using $1{,}000$ random seeds. For each shuffled dataset, we refit the embedding regressions and repeat validation-based model selection before evaluating the test predictions. We compare the observed standalone $R^2$ and incremental loss improvement with their respective shuffled distributions. For each statistic, the one-sided permutation $p$-value is $(c+1)/(n+1)$, where $n=1{,}000$ and $c$ is the number of shuffled values greater than or equal to the observed value.

\subsection{Comparisons across Time Periods and Coverage Groups}
\label{sec:method-decomp}
Two further cuts sharpen the interpretation. First, we split the test period into 2015--2019 and 2020--2022 and bootstrap the \emph{difference} $\Delta_{\text{pre}}-\Delta_{\text{post}}$ directly, using a single joint resample of all years so that the difference receives a valid confidence interval.

Second, we divide the session-window test sample into \emph{previously covered} and \emph{newly covered} stock-days. Previously covered stock-days also appear in the same-day panel. Newly covered stock-days are absent from that panel and enter only when after-hours, weekend, or holiday posts are included. These labels describe coverage under the two information-window rules, not different time periods or necessarily different firms. Both groups use session-window text and predictions from models trained on the full session-window panel, rather than predictions from the separately estimated same-day models. We call the paired difference in increments between the two groups, previously minus newly covered, the \emph{coverage contrast}.

\section{Results}
\label{sec:results}
All results are for the session-window panel unless noted, and the same-day panel is shown for comparison. Unless otherwise stated, we report incremental performance as the percentage reduction in equal-date MSE relative to HAR (Eq.~\eqref{eq:delta}). We report two-sided $95\%$ confidence intervals obtained from a moving-block bootstrap with a block length of five trading days. For firm-adjusted comparisons (Table~\ref{tab:firm}, Panel~B), the benchmark is HAR+firm rather than HAR.

\subsection{Standalone Predictability and Firm Identity}
\label{sec:res-standalone}

\begin{figure}[t]
  \centering
  \includegraphics[width=\linewidth]{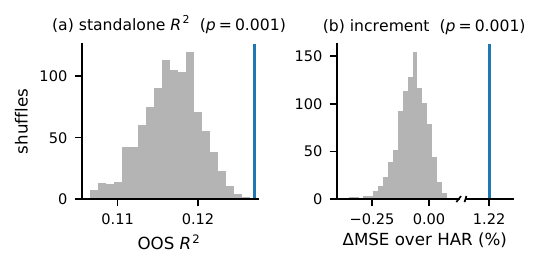}
\caption{Within-firm-year shuffle (session-window panel, seen firms). The observed standalone $R^2$ (a) sits just above its shuffle null, whereas the observed increment over HAR (b) lies above all shuffled counterparts.}
  \Description{Two null-distribution histograms with the observed statistic marked. The standalone statistic sits just above the null, while the increment lies far above it.}
  \label{fig:shuffle}
\end{figure}

An embedding-only model attains an out-of-sample $R^2$ of $0.130$ overall and $0.127$ on firms seen in training (Table~\ref{tab:firm}, Panel~A). On the same seen-firm sample, a regularized \texttt{permno} one-hot baseline using no text achieves $R^2=0.323$. Shuffling embeddings within firm-years preserves most of the standalone fit. Across $1{,}000$ shuffles, the median $R^2$ is $0.116$ overall and $0.117$ on seen firms, retaining $89\%$ and $92\%$ of the respective observed values. The observed $R^2$ nevertheless exceeds every shuffled value in both scopes ($p=0.001$, Figure~\ref{fig:shuffle}a), indicating that day-specific alignment also contributes to standalone performance. One might worry that the shuffle barely changes the embeddings, for example because a firm's posts are similar from day to day within a year. In that case the standalone $R^2$ would survive the shuffle even if it depended on day-specific text. However, the same shuffled datasets largely eliminate the embedding's gain over HAR (Section~\ref{sec:res-notidentity}, Figure~\ref{fig:shuffle}b), showing that the shuffle disrupts the alignment supporting the measured increment. Taken together, the one-hot baseline and the shuffle indicate that standalone predictability mainly reflects persistent firm or firm-year information, with a smaller component tied to the original day-to-text pairing. The shuffle is a diagnostic comparison, not an exact decomposition into these two sources, and a high standalone fit by itself is therefore not evidence that the text carries strong daily predictive information.

\begin{table}[t]
\caption{Firm-identity decomposition (session-window panel). Panel~A: standalone OOS $R^2$ against text-free firm baselines, with the share retained under the within-firm-year shuffle (shuffled median over observed) and the plus-one permutation $p$. Panel~B: equal-date $\Delta$MSE $\%$ of HAR${+}$firm${+}$embedding over HAR${+}$firm (block $=5$, two-sided $95\%$ CI) by subsample, with the shuffle $p$ of the HAR to HAR${+}$embedding increment. The coverage contrast is a paired bootstrap on the $1{,}971$ dates where both groups are present. For newly covered stock-days, all shuffled HAR to HAR${+}$embedding increments are at least as large as the observed increment ($p=1.000$).}
  \label{tab:firm}
  \footnotesize
  \setlength{\tabcolsep}{3.5pt}
  \begin{tabular}{lccccc}
    \toprule
    \multicolumn{6}{l}{\textbf{Panel A. Standalone predictability and firm identity}}\\
    Scope & Emb.\ $R^2$ & Firm-mean $R^2$ & One-hot $R^2$ & Retained & $p$ \\
    \midrule
    All  & $0.130$ & $0.127$ & $0.161$ & $89.2\%$ & $0.001$ \\
    Seen & $0.127$ & $0.276$ & $0.323$ & $92.1\%$ & $0.001$ \\
    \bottomrule
  \end{tabular}

  \vspace{4pt}
  \begin{tabular}{lcc}
    \toprule
    \multicolumn{3}{l}{\textbf{Panel B. Incremental value over HAR${+}$firm}}\\
    Subsample & $\Delta$MSE (\%) [$95\%$ CI] & $p$ \\
    \midrule
    All                 & $+1.19\ [0.92,1.46]$ & $0.001$ \\
    Seen                & $+1.18\ [0.84,1.52]$ & $0.001$ \\
    \quad Firm-selected windows (7/8) & $+1.06\ [0.74,1.39]$ & --- \\
    Unseen              & $+1.26\ [0.87,1.65]$ & $0.001$ \\
    Pre (2015--19)      & $+1.69\ [1.31,2.06]$ & $0.001$ \\
    Post (2020--22)     & $+0.10\ [-0.17,0.35]$ & $0.001$ \\
    Previously covered  & $+1.80\ [1.48,2.12]$ & $0.001$ \\
    Newly covered       & $-0.75\ [-1.19,-0.32]$ & $1.000$ \\
    Coverage contrast (paired) & $+2.55\ [2.02,3.08]$ & --- \\
    \bottomrule
  \end{tabular}
\end{table}

\subsection{Incremental Predictive Value over HAR}
\label{sec:res-increment}
Against the HAR benchmark, the incremental gain is small but statistically significant. On the session-window panel, adding embeddings improves the equal-date MSE by $+1.22\%$ $[0.94,1.49]$, with a QLIKE improvement of $+5.5\%$ (Table~\ref{tab:main}). The two gains are not on a common scale: MSE is evaluated on log realized variance, whereas QLIKE is evaluated on variance levels and depends on the ratio of realized to forecast variance, so the two percentage improvements cannot be compared directly. The interval of the MSE gain excludes zero. The consecutive-session subsample agrees closely, while the same-day panel shows a larger gain of $+1.83\%$, which Section~\ref{sec:res-coverage} decomposes. In RMSE terms, the gain is about $+0.61\%$.

\begin{table}[t]
\caption{Incremental value of embeddings over HAR (equal-date relative MSE improvement $\Delta$ of Eq.~\eqref{eq:delta}, block $=5$, two-sided $95\%$ CI).}
  \label{tab:main}
  \footnotesize
  \setlength{\tabcolsep}{3.5pt}
  \begin{tabular}{lccc}
    \toprule
    Panel & $\Delta$MSE (\%) & QLIKE (\%) & standalone $R^2$ \\
    \midrule
    Same-day      & $+1.83\ [1.38,2.26]$ & $+8.30\ [7.07,9.49]$ & $0.144$ \\
    Session-window        & $+1.22\ [0.94,1.49]$ & $+5.47\ [4.59,6.34]$ & $0.130$ \\
    \quad consecutive-session     & $+1.24\ [0.96,1.52]$ & $+5.22\ [4.37,6.11]$ & $0.129$ \\
    \bottomrule
  \end{tabular}
\end{table}

\subsection{Firm Controls and Day-Specific Alignment}
\label{sec:res-notidentity}

Section~\ref{sec:res-standalone} provided evidence that most standalone fit reflects firm or firm-year information. Because HAR already conditions on each firm's own lagged realized variance, that information does not automatically translate into an increment over HAR. Two complementary checks show that the increment survives the firm-effect control and that it disappears once the pairing between each day's text and its forecast target is broken.

First, adding embeddings to HAR${+}$firm reduces the equal-date MSE by $+1.18\%$ $[0.84,1.52]$ on seen firms (Table~\ref{tab:firm}, Panel~B). Validation, which may also select no firm adjustment at all, chooses a nonzero firm effect in seven of the eight rolling windows. In the 2015 test window it selects the null, so the comparison there reduces to HAR versus HAR${+}$embedding. Restricting to the seven windows with a selected firm effect, the firm-adjusted increment remains $+1.06\%$ $[0.74,1.39]$ on seen firms and $+1.13\%$ $[0.84,1.41]$ overall. On unseen firms, where the firm effect is zero by construction, the improvement is $+1.26\%$.

Second, shuffling the day-to-text pairing largely eliminates the HAR to HAR${+}$embedding increment, the comparison without the firm effect. For firms seen in training, none of the $1{,}000$ shuffled increments reaches the observed $+1.22\%$ (median $-0.07\%$, $p=0.001$, Figure~\ref{fig:shuffle}b). Unlike most of the standalone fit, the incremental gain therefore depends on day-specific alignment. This dependence does not by itself show that the text contains forward-looking information, since posts may instead describe contemporaneous events or market conditions that predict subsequent volatility (Section~\ref{sec:discussion}).

\subsection{Variation across Time Periods}
\label{sec:res-temporal}
\begin{figure}[t]
  \centering
  \includegraphics[width=\linewidth]{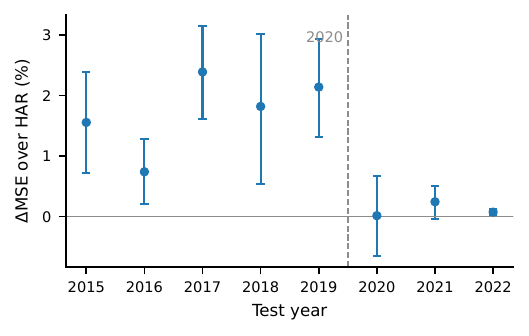}
\caption{Yearly equal-date MSE improvement over HAR (session-window panel, block $=5$). Positive and significant in each year of 2015--2019, small in 2020--2022.}
  \Description{Point estimates with error bars of the yearly MSE improvement, positive in 2015--2019 and small in 2020--2022.}
  \label{fig:yearly}
\end{figure}

The previous section established a small, alignment-dependent increment. We now ask whether it is stable over time. The pooled estimate masks substantial temporal heterogeneity. Figure~\ref{fig:yearly} shows annual increments: they are positive with intervals excluding zero in each of 2015 to 2019, whereas the 2020 to 2022 estimates are $+0.01\%$, $+0.24\%$, and $+0.07\%$, and only the interval for 2022 narrowly excludes zero. This pattern motivates an exploratory comparison of 2015--2019 with 2020--2022. Because the split point is chosen after inspecting these yearly estimates, the comparison describes heterogeneity. With this split, the improvement is $\Delta_{\text{pre}}=+1.72\%$ $[1.34,2.09]$ in 2015--2019 but $\Delta_{\text{post}}=+0.10\%$ $[-0.19,0.37]$ in 2020--2022. A direct bootstrap of the difference gives $\Delta_{\text{pre}}-\Delta_{\text{post}}=1.61$ percentage points $[1.15,2.09]$, excluding zero (Table~\ref{tab:period}). The decline is also visible in absolute MSE reductions ($0.016$ in 2015--2019 against $0.001$ in 2020--2022), so it is not an artifact of the change in the denominator of Eq.~\eqref{eq:delta}. After adding the firm effect and restricting to firms seen in training, the difference is $1.55$ percentage points $[1.00,2.11]$. In 2020--2022, the observed increment of $+0.10\%$ exceeds all $1{,}000$ shuffled values ($p=0.001$), although its bootstrap interval includes zero. These findings address different questions: the bootstrap interval quantifies uncertainty in the increment over HAR, whereas the permutation test compares the observed increment with those obtained after disrupting day-to-text alignment. The latter provides evidence of alignment-dependent information, but does not establish a positive expected improvement over HAR. We do not attribute the change to COVID-19 or to the meme-stock episode.

\begin{table}[t]
  \caption{Temporal decomposition and coverage interaction (equal-date $\Delta$MSE $\%$, block $=5$). Panel~A: HAR to HAR${+}$embedding period improvements and their direct bootstrapped difference (two-sided $95\%$ CI). The firm-adjusted row uses HAR${+}$firm as the benchmark on firms seen in training.
 Panel~B: HAR to HAR${+}$embedding increments by coverage $\times$ period with two-sided $95\%$ CIs from the same date-block bootstrap, and the paired coverage contrast (previously minus newly covered) within each period (exploratory).}
  \label{tab:period}
  \footnotesize
  \setlength{\tabcolsep}{1.5pt}
  \begin{tabular}{lccc}
    \toprule
    \multicolumn{4}{l}{\textbf{Panel A. Period difference, HAR to HAR${+}$embedding}}\\
    Panel & $\Delta_{\text{pre}}$ & $\Delta_{\text{post}}$ & $\Delta_{\text{pre}}-\Delta_{\text{post}}$ \\
    \midrule
    Same-day        & $+2.67\ [2.08,3.26]$ & $-0.14\ [-0.57,0.27]$ & $+2.81\ [2.08,3.54]$ \\
    Session-window      & $+1.72\ [1.34,2.09]$ & $+0.10\ [-0.19,0.37]$ & $+1.61\ [1.15,2.09]$ \\
    \quad Consec.& $+1.73\ [1.35,2.10]$ & $+0.15\ [-0.15,0.42]$ & $+1.58\ [1.11,2.07]$ \\
    \quad Firm-adj.\ (seen) & $+1.67\ [1.20,2.14]$ & $+0.12\ [-0.17,0.40]$ & $+1.55\ [1.00,2.11]$ \\
    \midrule
    \multicolumn{4}{l}{\textbf{Panel B. Coverage $\times$ period (two-sided $95\%$ CI)}}\\
    & Pre (2015--19) & Post (2020--22) & \\
    \midrule
    Previously covered & $+2.47\ [2.02,2.93]$ & $+0.38\ [0.12,0.64]$ & \\
    Newly covered    & $-0.84\ [-1.47,-0.22]$ & $-0.62\ [-1.04,-0.20]$ & \\
    Coverage contrast & $+3.33\ [2.56,4.11]$ & $+1.00\ [0.62,1.36]$ & \\
    \bottomrule
  \end{tabular}
\end{table}

\subsection{Does Broader Text Coverage Add Predictive Value?}
\label{sec:res-coverage}

We next test whether expanding textual coverage adds incremental information or merely changes the composition. The session-window rule enlarges the panel by $48\%$, almost entirely through off-hours and weekend content. This extra coverage does not translate into extra forecasting value. We divide the test stock-days into the previously and newly covered groups defined in Section~\ref{sec:method-decomp}. The firm-adjusted increment is $+1.80\%$ $[1.48,2.12]$ on previously covered stock-days but $-0.75\%$ $[-1.19,-0.32]$ on newly covered ones (Table~\ref{tab:firm}, Panel~B). We then bootstrap the coverage contrast directly, pairing the two groups by date and using only the dates on which both groups are present. The coverage contrast is $2.55$ percentage points $[2.02,3.08]$ after firm adjustment and $2.60$ percentage points $[2.07,3.13]$ for HAR to HAR${+}$embedding. Restricting the firm-adjusted comparison to seen firms gives $2.76$ percentage points $[2.13,3.41]$. All three intervals exclude zero. Across specifications, the increments on newly covered stock-days are negative or indistinguishable from zero, while the coverage contrast is consistently positive.

We next examine how the increment on newly covered stock-days varies across weekdays. The negative contribution is concentrated at the end of the week. By weekday, the HAR to HAR${+}$em\-bed\-ding increment on these stock-days is $-2.48\%$ $[-3.39,-1.60]$ for Friday stock-days and $-1.57\%$ $[-2.43,-0.71]$ for Thursday stock-days, whereas the intervals for Monday to Wednesday stock-days all contain zero. A Friday stock-day, $14\%$ of the newly covered group, typically pairs posts from the session window ending at Friday's close with Monday's realized variance. This variance includes the overnight return from Friday's close to Monday's open and can therefore reflect weekend news unavailable when the posts were written. This mismatch could explain the Friday figure but not the Thursday one, since a Thursday stock-day targets Friday's realized variance, which spans no weekend. For Monday forecast dates, whose information window includes weekend posts and whose target is typically Tuesday's realized variance, the increment is $-0.30\%$ $[-1.16,0.50]$ and is not statistically distinguishable from zero.

One possible explanation for the coverage contrast is that newly covered stock-days are more prevalent in the later period, when the overall increment is smaller. To assess this possibility, we compare previously and newly covered stock-days separately within 2015--2019 and 2020--2022 (Table~\ref{tab:period}, Panel~B). These comparisons measure the embedding increment over HAR without firm adjustment. The paired coverage contrast is positive in both periods, at $3.33$ percentage points $[2.56,4.11]$ in 2015--2019 and $1.00$ percentage points $[0.62,1.36]$ in 2020--2022, and the increment on newly covered stock-days is negative in both periods ($-0.84\%$ $[-1.47,-0.22]$ and $-0.62\%$ $[-1.04,-0.20]$). The positive contrast within both periods indicates that the coverage difference is not solely explained by the pre/post composition of the two groups.

The same pattern appears at the panel level. The HAR to HAR${+}$em\-bed\-ding gain of $+1.83\%$ on the same-day panel falls to $+1.22\%$ on the session-window panel. The smaller panel-level gain is consistent with the negative increments observed on newly covered stock-days. In short, more mentions do not imply more incremental forecasting information.

\subsection{Robustness Checks}
\label{sec:res-robust}
The checks below assess sensitivity to data construction, model specification, and evaluation choices. Table~\ref{tab:encoder} reports the encoder and link variants.

\begin{table*}[t]
\caption{Robustness to the encoder configuration (Panel~A, main panel) and to the post-to-firm link (Panel~B). Equal-date $\Delta$MSE $\%$ over HAR, block $=5$, two-sided $95\%$ CI. ``Retained'': shuffled median of the standalone $R^2$ over the observed value ($1{,}000$ shuffles, all test rows, and for Panel~B also seen firms). ``Pre'' and ``Post'': increments in 2015--2019 and 2020--2022. ``Coverage'': paired coverage contrast, previously minus newly covered, in percentage points. Values above $100\%$ indicate that the shuffled median standalone $R^2$ exceeds the observed value.}
  \label{tab:encoder}\label{tab:link}
  \footnotesize
  \setlength{\tabcolsep}{4pt}
  \begin{tabular}{lrcccccc}
    \toprule
    & Stock-days & standalone $R^2$ & retained & $\Delta$MSE (\%) & Pre & Post & Coverage \\
    \midrule
    \multicolumn{8}{l}{\textbf{Panel A. Encoder configuration}}\\
    Main: Llama-3.1-8B, layer 32 after norm, mean pooling, $192$ tokens & $202{,}145$ & $0.130$ & $89.2\%$ & $+1.22\ [0.94,1.49]$ & $+1.72$ & $+0.10$ & $+2.60$ \\
    Context cap $4{,}096$ tokens & $202{,}145$ & $0.141$ & $91.4\%$ & $+1.04\ [0.73,1.34]$ & $+1.75$ & $-0.54$ & $+3.05$ \\
    Layer 16 & $202{,}145$ & $0.027$ & $139.2\%$ & $+0.95\ [0.58,1.32]$ & $+1.48$ & $-0.25$ & $+2.42$ \\
    Layer 24 & $202{,}145$ & $0.057$ & $92.1\%$ & $+1.38\ [1.05,1.72]$ & $+2.03$ & $-0.06$ & $+3.04$ \\
    Layer 32 before norm & $202{,}145$ & $0.123$ & $89.6\%$ & $+1.17\ [0.90,1.43]$ & $+1.66$ & $+0.09$ & $+2.48$ \\
    Llama-2-7B, last-token pooling & $202{,}145$ & $0.051$ & $69.3\%$ & $+0.31\ [0.05,0.57]$ & $+0.39$ & $+0.13$ & $+1.74$ \\
    Ticker strings masked & $202{,}145$ & $-0.005$ & --- & $+1.24\ [0.83,1.64]$ & $+1.73$ & $+0.16$ & $+3.15$ \\
    \midrule
    \multicolumn{8}{l}{\textbf{Panel B. Post-to-firm link (retained: all / seen)}}\\

    Static link (earliest \texttt{permno} per string) & $156{,}503$ & $0.131$ & $96.0\%$ / $99.1\%$ & $+0.95\ [0.70,1.20]$ & $+1.32$ & $+0.12$ & $+1.99$ \\

    Date-valid link, no mention rule & $263{,}997$ & $0.093$ & $93.9\%$ / $95.8\%$ & $+0.94\ [0.72,1.16]$ & $+1.32$ & $+0.13$ & $+2.09$ \\

    \quad threshold $2\%$ & $214{,}293$ & $0.112$ & $89.8\%$ / $92.9\%$ & $+1.16\ [0.89,1.42]$ & $+1.65$ & $+0.11$ & $+2.54$ \\

    \quad threshold $5\%$ (main) & $202{,}145$ & $0.130$ & $89.2\%$ / $92.1\%$ & $+1.22\ [0.94,1.49]$ & $+1.72$ & $+0.10$ & $+2.60$ \\

    \quad threshold $10\%$ & $191{,}436$ & $0.140$ & $87.5\%$ / $90.3\%$ & $+1.28\ [0.99,1.56]$ & $+1.79$ & $+0.14$ & $+2.74$ \\

    \quad cashtags only & $71{,}518$ & $0.097$ & $92.3\%$ / $95.7\%$ & $+0.62\ [0.30,0.93]$ & $+0.79$ & $+0.24$ & $+2.33$ \\
    \bottomrule
  \end{tabular}
\end{table*}

\paragraph{Next-session targets.} For $0.9\%$ of stock-days the target is not the immediately following session (Section~\ref{sec:data-window}). Repeating the main and period comparisons on the consecutive-session subsample gives similar results (Tables~\ref{tab:main} and~\ref{tab:period}).

\paragraph{Evaluation and inference.} We recompute the bootstrap intervals with block lengths of $1$, $10$, and $22$ trading days in place of $5$, replace the equal-date estimand by the pooled stock-day estimand, and replace MSE by QLIKE. The HAR to HAR${+}$embedding increment, the period difference, and the coverage contrast keep intervals that exclude zero at every block length, the increment under the pooled estimand is $+0.69\%$ $[0.41,0.96]$, and QLIKE moves in the same direction as MSE (Table~\ref{tab:main}).

\paragraph{Post counts.} We fit a model with post counts only, and we add raw and unique post counts and the duplicate ratio to HAR. The count-only model has $R^2=-0.12$, and the counts add no stable improvement to HAR. These count-based features do not reproduce the predictive gain from embeddings.

\paragraph{Context length and layer.} A cap of $4{,}096$ tokens instead of $192$ raises the standalone $R^2$ from $0.130$ to $0.141$ but does not raise the increment ($+1.04\%$ $[0.73,1.34]$ against $+1.22\%$ at the main cap), and the standalone fit survives the shuffle (retained $91\%$), consistent with firm or firm-year information. Under the longer cap, the 2020--2022 increment is significantly negative ($-0.54\%$ $[-0.96,-0.15]$). Pooling the token representations of layer 16 or layer 24 instead of the final layer 32 (Section~\ref{sec:data-text}) gives much lower standalone $R^2$ ($0.027$ and $0.057$) but increments of similar size ($+0.95\%$ and $+1.38\%$). Using layer-32 representations before the model's final normalization yields an increment of $+1.17\%$, compared with $+1.22\%$ in the main specification.

\paragraph{Alternative encoder.} To check that the pattern is not specific to Llama-3.1, we embed the same texts with Llama-2-7B, using the last-token representation instead of mean pooling, and rerun the full pipeline on the same panel. The weaker embedding attains a standalone $R^2$ of only $0.051$ and a smaller increment of $+0.31\%$ $[0.05,0.57]$, but the alignment and coverage findings persist: none of the $1{,}000$ shuffles reaches the observed increment ($p=0.001$), and the coverage contrast remains positive ($1.74$ percentage points). Its period difference ($0.26$ percentage points $[-0.15,0.67]$) is not distinguishable from zero. Because this run changes the encoder and the pooling at once, it tests whether the pattern survives a different embedding, not which encoder is better.

\paragraph{Ticker masking.} Before embedding, we replace every ticker mention defined by the mention rule of Section~\ref{sec:data-text}, including cashtags and lower-case spellings of established tickers, with the word TICKER and rerun the pipeline on the same panel. Masking reduces standalone $R^2$ to $-0.005$, while the increment remains close to the unmasked estimate ($+1.24\%$ $[0.83,1.64]$ against $+1.22\%$). This contrast indicates that explicit ticker tokens are important for standalone performance but are not necessary for the measured increment.

\paragraph{Matching posts to firms.} We rerun the full pipeline under a static ticker table that keeps the earliest \texttt{permno} of each string, under the date-valid link with no mention rule and with thresholds of $2\%$, $5\%$, and $10\%$, and on cashtag mentions only. The static link gives an increment of $+0.95\%$ $[0.70,1.20]$, with the period difference and coverage contrast remaining positive. From no rule through the $10\%$ threshold, the increment grows from $+0.94\%$ to $+1.28\%$ while the shuffled share of the standalone fit falls from $93.9\%$ to $87.5\%$, consistent with mis-linked stock-days inflating the firm-identity share. The cashtag-only panel, about a third the size of the main panel, gives $+0.62\%$ $[0.30,0.93]$.

\paragraph{Within-day embedding dispersion.} Averaging the posts of a stock-day discards how much they disagree, and disagreement itself might predict volatility or differ between the two coverage groups. We therefore add a dispersion feature to the benchmark: one minus the norm of the stock-day mean of the unit-length embeddings, which is zero when all post embeddings are identical, including the $68\%$ of stock-days with a single unique text. On its own, it adds $+0.05\%$ $[-0.00,0.10]$ over HAR, where $-0.00$ denotes a negative lower bound rounded to zero. With post counts and dispersion in the benchmark, the embedding increment is $+0.87\%$ $[0.61,1.11]$ and the coverage contrast is $1.99$ percentage points $[1.52,2.46]$, so both the increment and the coverage contrast remain positive after controlling for post counts and embedding dispersion.

\section{Discussion and Limitations}
\label{sec:discussion}

\paragraph{What the increment is, and is not.}
Our tests show that the embedding increment survives firm-effect controls and depends on the original pairing between daily text and its forecast target. These findings identify predictive information beyond the specified benchmark, but do not distinguish among possible sources such as event descriptions, investor attention, or preference and belief.

\paragraph{Interpreting the temporal pattern.}
The temporal results suggest that the incremental value of WSB embeddings should not be assumed to persist across periods. Under the main specification, the gain is smaller in 2020--2022 than in 2015--2019, and the 2020--2022 estimate is not statistically distinguishable from zero. Changes in forum composition, volatility dynamics, or the speed at which markets incorporate information could contribute to this pattern, but our analysis does not distinguish among these explanations. The temporal pattern appears under every Llama-3.1 configuration in Table~\ref{tab:encoder}. With Llama-2 embeddings the point estimates move in the same direction ($+0.39\%$ in 2015--2019 and $+0.13\%$ in 2020--2022), but both increments are too small for their difference to be distinguishable from zero, so the temporal pattern is established for one encoder family and only weakly corroborated by a weaker one.

\paragraph{Scope of the sample and of the claims.} Our findings concern WSB-mentioned U.S. stock-days, evaluated with equal-date weighting at the horizon of the next eligible realized-variance observation. Nothing here speaks to the unconditional cross-section, to longer horizons, or to other platforms.

\paragraph{Modeling limitations.}
The augmentation is linear: ridge on frozen embeddings, with HAR coefficients fixed by the first stage. This restriction makes the increment interpretable, but joint estimation or nonlinear heads could yield different results, and we did not fit nonlinear models on the session-window panel. We report one encoder family in the main text and vary its context length, layer and model version in Section~\ref{sec:res-robust}. Larger or domain-adapted models could carry more information. The look-ahead concern of Section~\ref{sec:method-rolling} remains. The encoders' pretraining data may include information unavailable at historical forecast dates, so the reported gains should not be interpreted as real-time forecasting performance. This limitation does not prevent the diagnostic comparisons from distinguishing standalone fit from incremental value in the retrospective setting.

\paragraph{Data limitations.}
Deduplication merges only posts with identical text, so near-duplicates that differ by a few words remain separate posts, and their repeated content receives extra weight in the stock-day average. Date-valid linking and the mention rule reduce matching errors, but implicit references and company-name mentions may still be missed. The $5\%$ threshold was fixed before forecasting on the relinked panels, whereas the decision to adopt the date-valid panel rather than the static table as the main specification followed inspection of their results. Section~\ref{sec:res-robust} reports sensitivity to alternative linking and mention rules, including the static table.

\section{Conclusion}
\label{sec:conclusion}
An embedding-only model of Reddit posts achieves an out-of-sample $R^2$ of $0.13$ for log realized variance. Much of this fit reflects firm or firm-year information. A text-free firm baseline performs better than the embedding-only model on firms seen in training, and within-firm-year shuffling preserves most of the standalone fit. Masking ticker strings in the posts reduces standalone $R^2$ to $-0.005$. Adding embeddings to HAR reduces equal-date MSE by $1.22\%$. This small but statistically significant gain persists with firm controls and ticker masking, but is largely eliminated by within-firm-year shuffling. The increment is also temporally fragile: an exploratory period comparison shows a significant decline from 2015--2019 to 2020--2022, with the later-period increment not distinguishable from zero. Including off-hours content expands the panel by $48\%$, but embeddings provide no reliable forecasting improvement on the newly covered stock-days. More mentions therefore do not imply more incremental forecasting information. Together, these comparisons provide a diagnostic protocol that can be applied across text representations to assess their incremental value over strong benchmarks. It shifts the focus from standalone accuracy to what text adds beyond existing predictors, what information supports those gains, and whether they persist across periods and coverage groups.

\paragraph{Code and data availability.}
The code for the panel construction, the forecasting pipeline, and the diagnostics is not publicly released at the time of writing but is available from the author on request. The study uses Reddit posts from the Pushshift archive and TAQ and CRSP data licensed through WRDS, which cannot be redistributed.

\begin{acks}
I am deeply grateful to Ayan Bhattacharya, whose guidance shaped the framing and evaluation design of this paper. I also thank Lars Peter Hansen and Diana Petrova for their support, and gratefully acknowledge professional development funding from the Macro Finance Research Program at the University of Chicago.
\end{acks}

\bibliographystyle{ACM-Reference-Format}
\bibliography{references}


\begin{thebibliography}{32}


\ifx \showCODEN    \undefined \def \showCODEN     #1{\unskip}     \fi
\ifx \showISBNx    \undefined \def \showISBNx     #1{\unskip}     \fi
\ifx \showISBNxiii \undefined \def \showISBNxiii  #1{\unskip}     \fi
\ifx \showISSN     \undefined \def \showISSN      #1{\unskip}     \fi
\ifx \showLCCN     \undefined \def \showLCCN      #1{\unskip}     \fi
\ifx \shownote     \undefined \def \shownote      #1{#1}          \fi
\ifx \showarticletitle \undefined \def \showarticletitle #1{#1}   \fi
\ifx \showURL      \undefined \def \showURL       {\relax}        \fi
\providecommand\bibfield[2]{#2}
\providecommand\bibinfo[2]{#2}
\providecommand\natexlab[1]{#1}
\providecommand\showeprint[2][]{arXiv:#2}

\bibitem[Aggarwal et~al\mbox{.}(2024)]%
        {aggarwal2024meme}
\bibfield{author}{\bibinfo{person}{Dhruv Aggarwal}, \bibinfo{person}{Albert~H.
  Choi}, {and} \bibinfo{person}{Yoon-Ho~Alex Lee}.}
  \bibinfo{year}{2024}\natexlab{}.
\newblock \showarticletitle{The Meme Stock Frenzy: Origins and Implications}.
\newblock \bibinfo{journal}{\emph{Southern California Law Review}}
  \bibinfo{volume}{96}, \bibinfo{number}{6} (\bibinfo{year}{2024}),
  \bibinfo{pages}{1387--1415}.
\newblock


\bibitem[Alain and Bengio(2017)]%
        {alain2017understanding}
\bibfield{author}{\bibinfo{person}{Guillaume Alain} {and}
  \bibinfo{person}{Yoshua Bengio}.} \bibinfo{year}{2017}\natexlab{}.
\newblock \showarticletitle{Understanding Intermediate Layers Using Linear
  Classifier Probes}. In \bibinfo{booktitle}{\emph{International Conference on
  Learning Representations (ICLR), Workshop Track}}.
\newblock


\bibitem[Andersen et~al\mbox{.}(2003)]%
        {andersen2003modelling}
\bibfield{author}{\bibinfo{person}{Torben~G. Andersen}, \bibinfo{person}{Tim
  Bollerslev}, \bibinfo{person}{Francis~X. Diebold}, {and}
  \bibinfo{person}{Paul Labys}.} \bibinfo{year}{2003}\natexlab{}.
\newblock \showarticletitle{Modeling and Forecasting Realized Volatility}.
\newblock \bibinfo{journal}{\emph{Econometrica}} \bibinfo{volume}{71},
  \bibinfo{number}{2} (\bibinfo{year}{2003}), \bibinfo{pages}{579--625}.
\newblock


\bibitem[Atkins et~al\mbox{.}(2018)]%
        {atkins2018financial}
\bibfield{author}{\bibinfo{person}{Adam Atkins}, \bibinfo{person}{Mahesan
  Niranjan}, {and} \bibinfo{person}{Enrico Gerding}.}
  \bibinfo{year}{2018}\natexlab{}.
\newblock \showarticletitle{Financial News Predicts Stock Market Volatility
  Better than Close Price}.
\newblock \bibinfo{journal}{\emph{The Journal of Finance and Data Science}}
  \bibinfo{volume}{4}, \bibinfo{number}{2} (\bibinfo{year}{2018}),
  \bibinfo{pages}{120--137}.
\newblock
\href{https://doi.org/10.1016/j.jfds.2018.02.002}{doi:\nolinkurl{10.1016/j.jfds.2018.02.002}}


\bibitem[Baker et~al\mbox{.}(2016)]%
        {baker2016measuring}
\bibfield{author}{\bibinfo{person}{Scott~R. Baker}, \bibinfo{person}{Nicholas
  Bloom}, {and} \bibinfo{person}{Steven~J. Davis}.}
  \bibinfo{year}{2016}\natexlab{}.
\newblock \showarticletitle{Measuring Economic Policy Uncertainty}.
\newblock \bibinfo{journal}{\emph{The Quarterly Journal of Economics}}
  \bibinfo{volume}{131}, \bibinfo{number}{4} (\bibinfo{year}{2016}),
  \bibinfo{pages}{1593--1636}.
\newblock


\bibitem[Barndorff-Nielsen and Shephard(2002)]%
        {barndorffnielsen2002}
\bibfield{author}{\bibinfo{person}{Ole~E. Barndorff-Nielsen} {and}
  \bibinfo{person}{Neil Shephard}.} \bibinfo{year}{2002}\natexlab{}.
\newblock \showarticletitle{Econometric Analysis of Realised Volatility and Its
  Use in Estimating Stochastic Volatility Models}.
\newblock \bibinfo{journal}{\emph{Journal of the Royal Statistical Society:
  Series B}} \bibinfo{volume}{64}, \bibinfo{number}{2} (\bibinfo{year}{2002}),
  \bibinfo{pages}{253--280}.
\newblock


\bibitem[Baumgartner et~al\mbox{.}(2020)]%
        {baumgartner2020pushshift}
\bibfield{author}{\bibinfo{person}{Jason Baumgartner}, \bibinfo{person}{Savvas
  Zannettou}, \bibinfo{person}{Brian Keegan}, \bibinfo{person}{Megan Squire},
  {and} \bibinfo{person}{Jeremy Blackburn}.} \bibinfo{year}{2020}\natexlab{}.
\newblock \showarticletitle{The Pushshift Reddit Dataset}. In
  \bibinfo{booktitle}{\emph{Proceedings of the Fourteenth International AAAI
  Conference on Web and Social Media (ICWSM)}}, Vol.~\bibinfo{volume}{14}.
  \bibinfo{pages}{830--839}.
\newblock


\bibitem[Belinkov(2022)]%
        {belinkov2022probing}
\bibfield{author}{\bibinfo{person}{Yonatan Belinkov}.}
  \bibinfo{year}{2022}\natexlab{}.
\newblock \showarticletitle{Probing Classifiers: Promises, Shortcomings, and
  Advances}.
\newblock \bibinfo{journal}{\emph{Computational Linguistics}}
  \bibinfo{volume}{48}, \bibinfo{number}{1} (\bibinfo{year}{2022}),
  \bibinfo{pages}{207--219}.
\newblock


\bibitem[Bollen et~al\mbox{.}(2011)]%
        {bollen2011twitter}
\bibfield{author}{\bibinfo{person}{Johan Bollen}, \bibinfo{person}{Huina Mao},
  {and} \bibinfo{person}{Xiaojun Zeng}.} \bibinfo{year}{2011}\natexlab{}.
\newblock \showarticletitle{Twitter mood predicts the stock market}.
\newblock \bibinfo{journal}{\emph{Journal of Computational Science}}
  \bibinfo{volume}{2}, \bibinfo{number}{1} (\bibinfo{year}{2011}),
  \bibinfo{pages}{1--8}.
\newblock


\bibitem[Bradley et~al\mbox{.}(2024)]%
        {bradley2024place}
\bibfield{author}{\bibinfo{person}{Daniel Bradley}, \bibinfo{person}{Jan
  Hanousek, Jr.}, \bibinfo{person}{Russell Jame}, {and}
  \bibinfo{person}{Zicheng Xiao}.} \bibinfo{year}{2024}\natexlab{}.
\newblock \showarticletitle{Place Your Bets? The Value of Investment Research
  on Reddit's Wallstreetbets}.
\newblock \bibinfo{journal}{\emph{The Review of Financial Studies}}
  \bibinfo{volume}{37}, \bibinfo{number}{5} (\bibinfo{year}{2024}),
  \bibinfo{pages}{1409--1459}.
\newblock
\href{https://doi.org/10.1093/rfs/hhad098}{doi:\nolinkurl{10.1093/rfs/hhad098}}


\bibitem[Bybee et~al\mbox{.}(2024)]%
        {bybee2024business}
\bibfield{author}{\bibinfo{person}{Leland Bybee}, \bibinfo{person}{Bryan
  Kelly}, \bibinfo{person}{Asaf Manela}, {and} \bibinfo{person}{Dacheng Xiu}.}
  \bibinfo{year}{2024}\natexlab{}.
\newblock \showarticletitle{Business news and business cycles}.
\newblock \bibinfo{journal}{\emph{The Journal of Finance}}
  \bibinfo{volume}{79}, \bibinfo{number}{5} (\bibinfo{year}{2024}),
  \bibinfo{pages}{3105--3147}.
\newblock


\bibitem[Chen et~al\mbox{.}(2024)]%
        {chen2024expected}
\bibfield{author}{\bibinfo{person}{Yifei Chen}, \bibinfo{person}{Bryan Kelly},
  {and} \bibinfo{person}{Dacheng Xiu}.} \bibinfo{year}{2024}\natexlab{}.
\newblock \showarticletitle{Expected Returns and Large Language Models}.
\newblock \bibinfo{journal}{\emph{SSRN Electronic Journal}}
  (\bibinfo{year}{2024}).
\newblock
\shownote{Chicago Booth working paper, SSRN No.\ 4416687}.
\newblock


\bibitem[Clayton and Coppola(2025)]%
        {clayton2025faustian}
\bibfield{author}{\bibinfo{person}{Christopher Clayton} {and}
  \bibinfo{person}{Antonio Coppola}.} \bibinfo{year}{2025}\natexlab{}.
\newblock \bibinfo{title}{Financial Regulation and {AI}: A {F}austian Bargain?}
\newblock
\showeprint[arxiv]{2507.18747}


\bibitem[Cookson and Niessner(2020)]%
        {cookson2020w}
\bibfield{author}{\bibinfo{person}{J.~Anthony Cookson} {and}
  \bibinfo{person}{Marina Niessner}.} \bibinfo{year}{2020}\natexlab{}.
\newblock \showarticletitle{Why Don't We Agree? Evidence from a Social Network
  of Investors}.
\newblock \bibinfo{journal}{\emph{The Journal of Finance}}
  \bibinfo{volume}{75}, \bibinfo{number}{1} (\bibinfo{year}{2020}),
  \bibinfo{pages}{173--228}.
\newblock


\bibitem[Corsi(2009)]%
        {corsi2009simple}
\bibfield{author}{\bibinfo{person}{Fulvio Corsi}.}
  \bibinfo{year}{2009}\natexlab{}.
\newblock \showarticletitle{A simple approximate long-memory model of realized
  volatility}.
\newblock \bibinfo{journal}{\emph{Journal of Financial Econometrics}}
  \bibinfo{volume}{7}, \bibinfo{number}{2} (\bibinfo{year}{2009}),
  \bibinfo{pages}{174--196}.
\newblock


\bibitem[Da et~al\mbox{.}(2011)]%
        {da2011search}
\bibfield{author}{\bibinfo{person}{Zhi Da}, \bibinfo{person}{Joseph Engelberg},
  {and} \bibinfo{person}{Pengjie Gao}.} \bibinfo{year}{2011}\natexlab{}.
\newblock \showarticletitle{In search of attention}.
\newblock \bibinfo{journal}{\emph{The Journal of Finance}}
  \bibinfo{volume}{66}, \bibinfo{number}{5} (\bibinfo{year}{2011}),
  \bibinfo{pages}{1461--1499}.
\newblock


\bibitem[Dai(2024)]%
        {dai2024}
\bibfield{author}{\bibinfo{person}{Chaoxing Dai}.}
  \bibinfo{year}{2024}\natexlab{}.
\newblock \emph{\bibinfo{title}{Essays in Financial Econometrics}}.
\newblock \bibinfo{thesistype}{Ph.\,D. Dissertation}.
  \bibinfo{school}{University of Chicago}.
\newblock
\href{https://doi.org/10.6082/uchicago.12818}{doi:\nolinkurl{10.6082/uchicago.12818}}


\bibitem[Dubey et~al\mbox{.}(2024)]%
        {llama31}
\bibfield{author}{\bibinfo{person}{Abhimanyu Dubey} {et~al\mbox{.}}}
  \bibinfo{year}{2024}\natexlab{}.
\newblock \bibinfo{title}{The Llama 3 Herd of Models}.
\newblock
\shownote{arXiv:2407.21783}.
\newblock


\bibitem[Engle and Rangel(2008)]%
        {engle2008spline}
\bibfield{author}{\bibinfo{person}{Robert~F. Engle} {and}
  \bibinfo{person}{Jose~Gonzalo Rangel}.} \bibinfo{year}{2008}\natexlab{}.
\newblock \showarticletitle{The Spline-GARCH Model for Low-Frequency Volatility
  and Its Global Macroeconomic Causes}.
\newblock \bibinfo{journal}{\emph{The Review of Financial Studies}}
  \bibinfo{volume}{21}, \bibinfo{number}{3} (\bibinfo{year}{2008}),
  \bibinfo{pages}{1187--1222}.
\newblock


\bibitem[Gentzkow et~al\mbox{.}(2019)]%
        {GentzkowKellyTaddy2019TextAsData}
\bibfield{author}{\bibinfo{person}{Matthew Gentzkow}, \bibinfo{person}{Bryan
  Kelly}, {and} \bibinfo{person}{Matt Taddy}.} \bibinfo{year}{2019}\natexlab{}.
\newblock \showarticletitle{Text as Data}.
\newblock \bibinfo{journal}{\emph{Journal of Economic Literature}}
  \bibinfo{volume}{57}, \bibinfo{number}{3} (\bibinfo{year}{2019}),
  \bibinfo{pages}{535--574}.
\newblock
\href{https://doi.org/10.1257/jel.20181020}{doi:\nolinkurl{10.1257/jel.20181020}}


\bibitem[Ghysels et~al\mbox{.}(2006)]%
        {ghysels2006predicting}
\bibfield{author}{\bibinfo{person}{Eric Ghysels}, \bibinfo{person}{Pedro
  Santa-Clara}, {and} \bibinfo{person}{Rossen Valkanov}.}
  \bibinfo{year}{2006}\natexlab{}.
\newblock \showarticletitle{Predicting Volatility: Getting the Most out of
  Return Data Sampled at Different Frequencies}.
\newblock \bibinfo{journal}{\emph{Journal of Econometrics}}
  \bibinfo{volume}{131}, \bibinfo{number}{1--2} (\bibinfo{year}{2006}),
  \bibinfo{pages}{59--95}.
\newblock


\bibitem[Hansen and Lunde(2005)]%
        {hansen2005forecast}
\bibfield{author}{\bibinfo{person}{Peter~R. Hansen} {and}
  \bibinfo{person}{Asger Lunde}.} \bibinfo{year}{2005}\natexlab{}.
\newblock \showarticletitle{A Forecast Comparison of Volatility Models: Does
  Anything Beat a {GARCH}(1,1)?}
\newblock \bibinfo{journal}{\emph{Journal of Applied Econometrics}}
  \bibinfo{volume}{20}, \bibinfo{number}{7} (\bibinfo{year}{2005}),
  \bibinfo{pages}{873--889}.
\newblock


\bibitem[Huang et~al\mbox{.}(2023)]%
        {huang2023finbert}
\bibfield{author}{\bibinfo{person}{Allen~H. Huang}, \bibinfo{person}{Hui Wang},
  {and} \bibinfo{person}{Yi Yang}.} \bibinfo{year}{2023}\natexlab{}.
\newblock \showarticletitle{FinBERT: A Large Language Model for Extracting
  Information from Financial Text}.
\newblock \bibinfo{journal}{\emph{Contemporary Accounting Research}}
  \bibinfo{volume}{40}, \bibinfo{number}{2} (\bibinfo{year}{2023}),
  \bibinfo{pages}{806--841}.
\newblock


\bibitem[Hutto and Gilbert(2014)]%
        {HuttoGilbert2014}
\bibfield{author}{\bibinfo{person}{Clayton~J. Hutto} {and}
  \bibinfo{person}{Eric Gilbert}.} \bibinfo{year}{2014}\natexlab{}.
\newblock \showarticletitle{{VADER:} A Parsimonious Rule-Based Model for
  Sentiment Analysis of Social Media Text}. In
  \bibinfo{booktitle}{\emph{Proceedings of the Eighth International Conference
  on Weblogs and Social Media (ICWSM 2014)}}. \bibinfo{pages}{216--225}.
\newblock


\bibitem[Kogan et~al\mbox{.}(2009)]%
        {kogan2009predicting}
\bibfield{author}{\bibinfo{person}{Shimon Kogan}, \bibinfo{person}{Dimitry
  Levin}, \bibinfo{person}{Bryan~R. Routledge}, \bibinfo{person}{Jacob~S.
  Sagi}, {and} \bibinfo{person}{Noah~A. Smith}.}
  \bibinfo{year}{2009}\natexlab{}.
\newblock \showarticletitle{Predicting Risk from Financial Reports with
  Regression}. In \bibinfo{booktitle}{\emph{Proceedings of Human Language
  Technologies: The 2009 Annual Conference of the North American Chapter of the
  Association for Computational Linguistics}}. \bibinfo{publisher}{Association
  for Computational Linguistics}, \bibinfo{address}{Boulder, Colorado},
  \bibinfo{pages}{272--280}.
\newblock
\urldef\tempurl%
\url{https://aclanthology.org/N09-1031/}
\showURL{%
\tempurl}


\bibitem[Loughran and McDonald(2011)]%
        {loughran2011liability}
\bibfield{author}{\bibinfo{person}{Tim Loughran} {and} \bibinfo{person}{Bill
  McDonald}.} \bibinfo{year}{2011}\natexlab{}.
\newblock \showarticletitle{When Is a Liability Not a Liability? Textual
  Analysis, Dictionaries, and 10-Ks}.
\newblock \bibinfo{journal}{\emph{The Journal of Finance}}
  \bibinfo{volume}{66}, \bibinfo{number}{1} (\bibinfo{year}{2011}),
  \bibinfo{pages}{35--65}.
\newblock


\bibitem[Manela and Moreira(2017)]%
        {manela2017news}
\bibfield{author}{\bibinfo{person}{Asaf Manela} {and} \bibinfo{person}{Alan
  Moreira}.} \bibinfo{year}{2017}\natexlab{}.
\newblock \showarticletitle{News Implied Volatility and Disaster Concerns}.
\newblock \bibinfo{journal}{\emph{Journal of Financial Economics}}
  \bibinfo{volume}{123}, \bibinfo{number}{1} (\bibinfo{year}{2017}),
  \bibinfo{pages}{137--162}.
\newblock


\bibitem[Mullainathan and Spiess(2017)]%
        {mullainathan2017machine}
\bibfield{author}{\bibinfo{person}{Sendhil Mullainathan} {and}
  \bibinfo{person}{Jann Spiess}.} \bibinfo{year}{2017}\natexlab{}.
\newblock \showarticletitle{Machine Learning: An Applied Econometric Approach}.
\newblock \bibinfo{journal}{\emph{Journal of Economic Perspectives}}
  \bibinfo{volume}{31}, \bibinfo{number}{2} (\bibinfo{year}{2017}),
  \bibinfo{pages}{87--106}.
\newblock
\href{https://doi.org/10.1257/jep.31.2.87}{doi:\nolinkurl{10.1257/jep.31.2.87}}


\bibitem[Ncume et~al\mbox{.}(2022)]%
        {ncume2022volatility}
\bibfield{author}{\bibinfo{person}{Vuyo Ncume}, \bibinfo{person}{T.~L. van
  Zyl}, {and} \bibinfo{person}{A. Paskaramoorthy}.}
  \bibinfo{year}{2022}\natexlab{}.
\newblock \showarticletitle{Volatility Forecasting Using Deep Learning and
  Sentiment Analysis}.
\newblock \bibinfo{journal}{\emph{arXiv preprint arXiv:2210.12464}}
  (\bibinfo{year}{2022}).
\newblock


\bibitem[Patton(2011)]%
        {patton2011volatility}
\bibfield{author}{\bibinfo{person}{Andrew~J. Patton}.}
  \bibinfo{year}{2011}\natexlab{}.
\newblock \showarticletitle{Volatility Forecast Comparison Using Imperfect
  Volatility Proxies}.
\newblock \bibinfo{journal}{\emph{Journal of Econometrics}}
  \bibinfo{volume}{160}, \bibinfo{number}{1} (\bibinfo{year}{2011}),
  \bibinfo{pages}{246--256}.
\newblock


\bibitem[Rahimikia et~al\mbox{.}(2021)]%
        {rahimikia2021realised}
\bibfield{author}{\bibinfo{person}{Eghbal Rahimikia}, \bibinfo{person}{Stefan
  Zohren}, {and} \bibinfo{person}{Ser-Huang Poon}.}
  \bibinfo{year}{2021}\natexlab{}.
\newblock \showarticletitle{Realised Volatility Forecasting: Machine Learning
  via Financial Word Embedding}.
\newblock \bibinfo{journal}{\emph{arXiv preprint arXiv:2108.00480}}
  (\bibinfo{year}{2021}).
\newblock


\bibitem[Ranco et~al\mbox{.}(2015)]%
        {ranco2015effects}
\bibfield{author}{\bibinfo{person}{Gabriele Ranco}, \bibinfo{person}{Darko
  Aleksovski}, \bibinfo{person}{Guido Caldarelli}, \bibinfo{person}{Miha Gr{\v
  c}ar}, {and} \bibinfo{person}{Igor Mozeti{\v c}}.}
  \bibinfo{year}{2015}\natexlab{}.
\newblock \showarticletitle{The Effects of Twitter Sentiment on Stock Price
  Returns}.
\newblock \bibinfo{journal}{\emph{PLOS ONE}} \bibinfo{volume}{10},
  \bibinfo{number}{9} (\bibinfo{year}{2015}), \bibinfo{pages}{e0138441}.
\newblock
\href{https://doi.org/10.1371/journal.pone.0138441}{doi:\nolinkurl{10.1371/journal.pone.0138441}}


\end{thebibliography}

\end{document}